\documentstyle[prb,aps,epsf,twocolumn]{revtex}
\begin{document} \draft

% Email address for Z. Tesanovic:	zbt@pha.jhu.edu

\title{Landau Levels and Quasiparticle Spectrum of Extreme
Type-II Superconductors}

\author{Zlatko Te{\v s}anovi{\' c}$^{a,}$\cite{zbt}
and 
Pedro D. Sacramento$^{a,b}$}
\address{$^a$Department of Physics and Astronomy, Johns Hopkins University,
Baltimore, MD 21218, USA\\
$^b$
Departamento de F\'{\i}sica and CFIF, Instituto
Superior T\'{e}cnico, Av. Rovisco Pais, 1096 Lisboa Codex, Portugal
\\ {\rm(\today)}}
\maketitle
\begin{abstract}
%\address{~
%\parbox{14cm}{\rm
%\medskip
At high magnetic fields and low temperatures, numerous extreme type-II
superconductors exhibit Landau quantization of electronic
motion. We present an analytic construction of the quasiparticle
spectrum in this regime, based on the high-field expansion. 
The spectrum is gapless and is separated from the familiar low-field
regime by a quantum level-crossing transition. Such low energy
excitations should lead to observable effects in thermodynamics
and transport.
%}}
\end{abstract}
%\maketitle
%pocetak
\narrowtext
Recent observations of the de Haas-van Alphen (dHvA) oscillations
in many extreme type-II superconducting 
materials\cite{exp} clearly indicate the presence
of Landau level (LL) quantization within a high-magnetic field ($H$), 
low-temperature ($T$) ``pocket", surrounding the $H_{c2}(0)$ point
in the $H-T$ phase diagram (Fig. 1). The size of this pocket
in conventional type-II systems (like Nb) is expected
to be negligible. This is so because, within the BCS
theory, the scale of the cyclotron splitting between LLs near
$H_{c2}(0)$, $\omega_c\sim\omega_{c2}(0)\equiv eH_{c2}(0)/m^*c$, is set by
the condensation energy, $\sim T_{c0}^2/E_F$, and
should be much smaller than either the thermal smearing, $\sim T$,
or the gap $\Delta (T,H)$, the scale for both being $T_{c0}$.
Additional smearing due to disorder, $\Gamma$, makes this tiny pocket
effectively irrelevant. In clean, intrinsically extreme type-II
systems, the situation is significantly different. We {\em define}
the extreme type-II systems as those with the slope of the
upper critical field at $T_{c0}$, $H_{c2}^{\prime}\geq 0.3$ 
Tesla/Kelvin.\cite{footi} In such systems 
$\omega_{c2}(0)$ is {\em comparable} to $T_{c0}$ and there is 
large pocket in the phase diagram
in which the LL structure {\em within} the superconducting
phase is well defined, i.e. $\omega_c > \Delta (T,H),T,\Gamma$.
Numerous superconductors belong to this extreme type-II family:
high-$T_c$ cuprates, A15's, 
boro-carbides, many organics, etc.
The boundaries of the ``extreme" pocket, $H^*$ and
$T^*$, defined by $\omega_c \sim\Delta (T=0,H)$ and
$\omega_c \sim T$, extend to $H$ as low as $H^*\sim 50\% H_{c2}(0)$
and $T$ as high as $T^*\sim 30\% T_{c0}$ (Fig. 1).\cite{exp}

Within this ``extreme" pocket the solution 
to the superconducting problem must fully incorporate the LL structure
of the normal state. This leads to a set
of Bogoliubov-de Gennes (BdG) equations for the LL quantized
quasiparticles in presence of the 
gap function $\Delta ({\bf r})$, 
describing the Abrikosov vortex lattice.\cite{sasa,norman}
There are two main sources of difficulty in solving these equations.
First, the number of LLs involved in the pairing is typically 40-100
and the matrix elements of the gap function between these states,
$\Delta_{nm}({\bf q})$, are in general quite complex. A more
serious difficulty, however, is that the basis which diagonalizes
the BCS Hamiltonian involves {\em combined} ``rotations"
{\em both} in the Nambu space {\em and} the 
LL basis. To illustrate why this 
fact seriously impedes analytic progress we first note that,
to the leading order in $\Delta/\omega_c \ll 1$, the quasiparticle
excitation spectrum {\em near} the Fermi surface (FS) can be found exactly: 
$E =\pm \sqrt{\varepsilon _{n}(k_z)^2 + |\Delta_{nn}({\bf q})|{^2}}$.\cite{sasa}
It is tempting to generalize this and 
conclude that the full solution takes the form:
$E =\pm \sqrt{\Sigma_{{\bar n}{\bar n}}^2 + 
|D_{{\bar n}{\bar n}}|^2}$, where 
$\Sigma_{{\bar n}{\bar n}}({\bf q},k_z)$ and
$D_{{\bar n}{\bar n}}({\bf q},k_z)$ are the normal and pairing
self-energies, respectively. The physical meaning of such 
solution would be that we first rotate the LL basis,
$\{n\}\to\{{\bar n}\}$, and then diagonally pair up these new
``normal" quasiparticles. This would
lead to a simple description, 
suitable for applications
of our standard theoretical machinery. Unfortunately, this
conclusion would be incorrect: beyond the leading order, the
full solution cannot be put in the above simple form as
the ``normal" and ``pairing" self-energies cannot be 
{\em simultaneously} diagonalized.\cite{footip} This leads to a
loss of physical transparency and makes it difficult
to interpret results in familiar terms.
Most of the work beyond 
leading order has been numerical.\cite{sasa,norman,bristol,bruun}

In this Letter we introduce an analytic approach which allows for
a transparent and systematic evaluation of 
corrections to the leading order results. The basic idea can 
be illustrated by an example of an interacting Fermi liquid: there 
various properties are in general
very difficult to evaluate. However, as long as we are interested
in the low $T$ thermodynamics and low energy transport,
these properties can be accurately described 
by the {\em quasiparticle} representation of the excitations
near the FS. This leads to the low $T$ 
thermodynamics which is in
one-to-one correspondence with that of non-interacting fermions,
but with a modified dispersion relation and a reduced spectral weight. 
In this paper, we construct such a ``quasiparticle"
representation of the present problem. Of course, there are
no interactions in our case but our aim is similar.
The original partition function is 
replaced by the one of ``quasiparticles"
whose energies {\em do have} the desired form 
$E =\pm \sqrt{\tilde\varepsilon_{nn}^2 + 
Z^2|\tilde\Delta_{nn}|^2}$, and which faithfully represents the
low $T$ (or low energy) properties of the
original problem.  Here  
$\tilde\varepsilon_{nn}({\bf q},k_z)$ and 
$\tilde\Delta_{nn}({\bf q},k_z)$ are the {\em renormalized} normal and pairing
self-energies and $Z$ is the "quasiparticle" renormalization
factor. All can be computed to the desired accuracy
in $\Delta/\omega_c$ from a general expression which we provide.
Using this ``quasiparticle" representation we derive the following
results: first, we evaluate the leading order corrections to
``normal" and ``pairing" self-energies and find that they are of order
$\sim \Delta^2/\omega_c$ and $\sim \Delta^3/\omega_c^2$, respectively.
Therefore, for $\Delta/\omega_c \ll 1$, the features of
the spectrum obtained in the leading order are only
moderately affected.  In particular, we show 
that the quasiparticle excitations remain
{\em gapless to all orders in perturbation theory} until
$\Delta/\omega_c$ exceeds certain critical value. As
$\Delta/\omega_c$ increases further ($H<H^*$),
the original LL structure is obliterated by
strong LL mixing induced by off-diagonal $\Delta_{nm}$'s. This 
exact result shows that the {\em gapless} 
``extreme" pocket is truly a novel
state of type-II superconductors and is separated from the
{\em gapped} low-field mixed phase by a quantum level crossing
transition. In this context, we point to an important difference
between 3D and 2D systems. We next present
our numerical results for the excitation
spectrum and find that they are in agreement
with the analytic results.
We hope to inspire experiments which could
test our results and provide novel insight
into the physics of high-field, extreme type-II
superconductors.

Our starting point are the BdG equations:
\begin{eqnarray} 
E
u_
{k_{z},{\bf q},n}^{N}
=\varepsilon_n(k_z)u_{k_{z},{\bf q},n}^{N}
+\sum_{m} \Delta_{nm}({\bf q})v_
{k_{z},{\bf q},m}^{N}
\nonumber \\
E
v_
{k_{z},{\bf q},n}^{N}
=-\varepsilon_n(k_z)v_{k_{z},{\bf q},n}^{N}
+\sum_{m} \Delta^{\ast}_{mn}({\bf q})u_{k_{z},{\bf q},m}^{N}~~, 
\label{ei}
\end{eqnarray}
where $\varepsilon_n(k_z)\equiv
\hbar^{2}k_{z}^{2}/2m
+n\hbar \omega _{c}-\mu$, $n$ is the LL index, $\hbar\omega_c /2$
has been absorbed into the chemical potential, $\mu$, ${\bf q}$
is the magnetic wavevector and $k_z$ is the wavevector along
the field direction.\cite{sasa}
$\Delta_{nm}({\bf q})=\sum_{j} 
\Delta_{n,m}^{j}({\bf q})$ is the matrix 
element of $\Delta({\bf r})=\sum_{j} 
\Delta_{j}({\bf r})$\cite{footii} between electronic
states $(k_{z},{\bf q},n)$ and $(-k_{z},-{\bf q},m)$.\cite{sasa}
$N$ is the number of LLs which
participate in pairing. The solution of the above equations
gives the quasiparticle states and the energy spectrum within
the superconducting phase. 

As one crosses into the superconducting state and $\Delta$ becomes
finite, there are both perturbative and non-perturbative effects
in Eq. (\ref{ei}). We first isolate the non-perturbative part by
noticing that the particle bands ($u$-bands) with the LL
index $n_1 = n+p$  and the hole bands
($v$-bands) with the LL index $n_2 = n-p$ ($p = 0,\pm 1,\pm 2,\dots$)
cross at energies $p\hbar\omega_c$ whenever 
$k_z = \pm k_{Fn}\equiv \sqrt{2m(\mu - n\hbar\omega_c)}$,
where $k_{Fn}$ are the Fermi wavevectors
of different quasi one-dimensional LL bands.
Additional band crossings take place at energies $(p+1/2)\hbar\omega_c$
and for $k_z = \pm \sqrt{k^2_{Fn}+ m\hbar\omega_c}$. 
At the band crossings near $k_z=\pm k_{Fn}$
the matrix elements $\Delta_{n+p,n-p}({\bf q})$ 
act non-perturbatively and must be treated by a 
degenerate perturbation theory. The same is true for
$\Delta_{n+p,n-1-p}({\bf q})$ 
near $k_z = \pm \sqrt{k^2_{Fn}+ m\hbar\omega_c}$. 
The effect of all other matrix elements, 
$\Delta_{n+p,n(-1)-p'}({\bf q})$ ($p\not= p'$),
is purely perturbative and can be computed by an expansion 
in $\Delta/\omega_c$. Near $H_{c2}$, where $\Delta/\omega_c \ll 1$,
that effect is small and we can obtain the analytic form for
the BCS excitation spectrum near $k_z=\pm k_{Fn}$:
\begin{equation}
E_{n,p}=p\hbar\omega_c\pm \sqrt{\varepsilon _{n}(k_z)^2 + 
|\Delta_{n+p,n-p}({\bf q})|{^2}}~~,
\label{eii}
\end{equation}
where $p=0,\pm 1,\pm 2,\dots$. A similar expression can be written
for the spectrum around 
$k_z \sim \pm \sqrt{k^2_{Fn}+ m\hbar\omega_c}$. Typically,
we are interested in the quasiparticle spectrum when we discuss
low $T$ (or low energy) properties,
{\it i.e.} $(T,\omega)\ll \Delta (T,H)<\omega_c$.  In that case
only the quasiparticles near the FS are important,
$k_z\sim\pm k_{Fn}$ and $p=0$, and it suffices to consider only
the $E_{n,p=0}=\pm \sqrt{\varepsilon _{n}(k_z)^2 + 
|\Delta_{nn}({\bf q})|{^2}}$ bands. 
The excitations from other, $p\not= 0$, bands (\ref{eii})
are gapped by energies which are too high, 
$E_{n,p\not= 0}(k_z\sim\pm k_{Fn})\sim p\hbar\omega_c$ and
their contribution to the quasiparticle thermodynamics is
negligible in the low $T$ ($T\ll \Delta (T,H)\ll\omega_c$)
regime.  This is the content of the
so-called ``diagonal approximation" (DA) for 
the excitation spectrum.\cite{sasa,footiii}

The main issue addressed in this paper is how to go beyond 
Eq. (\ref{eii}) analytically. As one moves further below
$H_{c2}$, $\Delta/\omega_c$ increases (while still $\ll 1$)
and the corrections coming from
$\Delta_{n+p,n-p'}({\bf q})$ ($p\not= p'$) must be included.
A {\em pedestrian} perturbation
theory, starting from (\ref{eii}), leads to expressions
whose physical content is difficult to analyze. The reason is
that (\ref{eii}) already involves
a full Nambu rotation. Consequently, all other
terms generated by the pedestrian perturbation theory
are fully Nambu rotated. This makes 
it impossible to disentangle the ``normal"
from the ``paring" part of the problem.\cite{footip} In contrast, our
standard theoretical machinery works best when we 
can neatly separate the normal from the pairing part.
We propose here a different
method of computing the excitation spectrum.
Consider the partition function 
specified by the solution of BdG equations
(\ref{ei}),
$Z = \prod_{\omega,{\bf q},k_z}
{\cal Z}(i\omega,{\bf q},k_z)$,
where
\begin{equation}
{\cal Z}=\prod_n\int d\psi_nd{\bar \psi}_n
\exp\bigl\{ \sum_{n,m}{\bar \psi}_n(i\omega{\cal I} - {\cal H})_{nm}
\psi_m\bigr\}~~.
\label{eiii}
\end{equation}
Here the set of Grassman variables
$\{u_n,v_n\}(i\omega,{\bf q},k_z)$ 
representing the particle (hole)
bands of (\ref{ei}) has been arranged in spinors
${\bar\psi}_n\equiv ({\bar u}_n~~{\bar v}_n)$. ${\cal I}$ and
${\cal H}({\bf q},k_z)$ are two $2N\times 2N$ matrices which
are, respectively, a unit matrix and a Hamiltonian
whose form is evident from the right hand side
of Eq. (\ref{ei}). $\{\omega\}$ is the set of fermionic
Matsubara frequencies.
Note that ${\cal Z}(i\omega,{\bf q},k_z)$ is a determinant of a matrix,
${\cal Z}=\det \Vert i\omega{\cal I} - {\cal H}\Vert $. When analytically 
continued, $E\to i\omega$, this is precisely the secular determinant
of the BdG equations (\ref{ei}). Thus, the solutions
for quasiparticle energies in (\ref{ei}) are the zeroes
of ${\cal Z}(i\omega\to E,{\bf q},k_z)$. 

Within our ``extreme"
pocket ($T\ll \Delta (T,H)\ll\omega_c$) the main contributions
to $Z$ come from $k_z\sim\pm k_{Fn}$ and $E_{n,p=0}$ bands. 
In order to evaluate ${\cal Z}(i\omega,{\bf q},k_z\sim k_{Fn})$
we first integrate out 
all $\psi_{n,p\not= 0}(i\omega,{\bf q},k_z\sim k_{Fn})$ in
(\ref{eiii}):
\begin{equation}
{\cal Z}={\cal Z}_{p\not= 0}
\int d\psi_{n,0}d{\bar \psi}_{n,0}
\exp\bigl\{ {\bar \psi}_{n,0}(i\omega{\cal I} - {\tilde{\cal H}})_{00}
\psi_{n,0}\bigr\}.
\label{eiv}
\end{equation}
${\cal Z}_{p\not= 0}(i\omega,{\bf q},k_z\sim k_{Fn})$ is what
one gets from (\ref{eiii}) with the $E_{n,p=0}$ band
{\em excluded}. It describes the contribution to
thermodynamics arising from all other bands, $E_{n,p\not= 0}$,
by themselves. Since $p\not= 0$ bands are far from the
FS, $E_{n,p\not= 0}({\bf q},k_z\sim k_{Fn})
\sim p\hbar\omega_c$, such contributions are activated with
a large gap, {\it i.e.} $\sim \exp (-p\hbar\omega_c/T)\ll 1$,
and their contribution is very small. Thus, the dominant
contribution comes from 
$\det\Vert (i\omega{\cal I} - {\tilde{\cal H}})_{00}\Vert $.
Here ${\cal I}_{00}$ is a $2\times 2$ unit matrix while 
${\tilde{\cal H}}_{00}\equiv {\cal H}^0 + {\cal H}'$ is
a $2\times 2$ Nambu matrix. The diagonal elements of
$\{{\cal H}^0,{\cal H}'\}(i\omega,{\bf q},k_z\sim k_{Fn})$ are 
$\{\pm\varepsilon_n(k_z),\pm\Sigma_{nn} (i\omega,{\bf q},k_z)\}$ 
while the off-diagonal ones are
$\{\Delta_{nn}({\bf q}),D_{nn}(i\omega,{\bf q},k_z)\}$ and 
their complex conjugates (c.c.). By itself, ${\cal H}^0$ determines
the spectrum of quasiparticles at the FS 
(described by $\psi_{n,p = 0}$) in the
DA (\ref{eii}). ${\cal H}'$ is a correction
arising from coupling to other spinor bands,
$\psi_{n,p\not= 0}$, with $\Sigma_{nn}$ and $D_{nn}$ being
the normal and pairing self-energies, respectively. Since
the $p\not= 0$ bands are separated from the FS
by $p\hbar\omega_c$, these self-energies contain denominators
$\sim p\hbar\omega_c$ and are small for $\Delta/\omega_c \ll 1$.
The explicit form is obtained from:
\begin{equation}
{\cal H}' = \sum_{p\not=0}\sum_{p'\not=0}
{\cal H}_{0p}\bigl(i\omega{\cal I}-{\cal H}\bigr)^{-1}_{pp'}
{\cal H}_{p'0}~~.
\label{ev}
\end{equation}
${\cal H}_{0p}$ is a $2\times2$ Nambu matrix whose diagonal
elements are zero and off-diagonal ones $\Delta_{n,n+p}({\bf q})$
and its c.c. The resolvent appearing in (\ref{ev}) is
that of the BdG system (\ref{ei}) for $k_z\sim k_{Fn}$ but
with the $p=0$ spinor band (\ref{eii}) excluded.

Up to this point there are no approximations and the expression
for ${\cal H}'$ is formally exact. The problem is that the
resolvent for the $p\not=0$ bands 
involves inversion of a large matrix. However,
for $\Delta\ll \omega_c$, the resolvent can be calculated
perturbatively, to a desired accuracy in $\Delta/\omega_c$.
In general, it is useful to first perform the Nambu rotation
for the $p\not= 0$ bands 
to get rid of the degeneracies between the $n+p$ particle
and $n-p$ hole bands and then proceed with the perturbative
expansion. It turns out that this is not necessary if we
are only after the leading order corrections. In this way
we obtain:
$$\Sigma_{nn} =
\sum_{p\not=0}\frac{|\Delta_{n,n+p}({\bf q})|^2}
{i\omega + \varepsilon_{n+p}(k_z)}~~;~~D_{nn} = $$
\begin{equation}
= \sum_{p\not=0,p'\not=0}
\frac{\Delta_{n,n+p}({\bf q})\Delta^*_{n+p,n+p'}(-{\bf q})
\Delta_{n+p',n}({\bf q})}{(i\omega + \varepsilon_{n+p}(k_z))
(i\omega - \varepsilon_{n+p'}(k_z))} .
\label{evi}
\end{equation}
$\Sigma_{nn}(i\omega,{\bf q},k_z)$ and
$D_{nn}(i\omega,{\bf q},k_z)$ are the normal and pairing
self-energies, respectively. They have a frequency dependence
arising from the coupling to the $p\not= 0$ spinor bands
through the matrix elements
$\Delta_{n,n+p}({\bf q})$. We can view them as self-energies
of some ``interacting" system and proceed to construct the
``quasiparticle" representation of (\ref{eiii}). This
construction uses a {\em Fermi surface expansion} of (\ref{evi}).
First, we determine the {\em renormalized} ``quasiparticle"
energies from $\tilde\varepsilon_{nn} ({\bf q},k_z)=
\varepsilon_n (k_z) + \Sigma_{nn}(i\omega=\tilde
\varepsilon_{nn} ({\bf q},k_z),{\bf q},k_z)$. To the leading order in
$\Delta/\omega_c$ this gives:
\begin{equation}
\tilde\varepsilon_{nn} ({\bf q},k_z)\approx
\varepsilon_n (k_z) + \sum_{p\not=0}\frac{|\Delta_{n,n+p}({\bf q})|^2}
{p\omega_c}~~.
\label{evii}
\end{equation}
We also evaluate the
``quasiparticle" renormalization factor, $Z_n^{-1}\equiv
1 - \partial\Sigma_{nn}/\partial i\omega |_{i\omega =\tilde\varepsilon_{nn}}$;
to the leading order $ Z_n \approx 1 - 
\sum_{p\not=0}|\Delta_{n,n+p}({\bf q})|^2/
p^2\omega_c^2$.
Similarly, we can compute the ``renormalized" shape of the
FS. The ``renormalized" Fermi velocity is given
by:
\begin{equation}
\frac{\tilde v_{Fn}({\bf q})}{v_{Fn}}\approx
1 - \sum_{p\not=0}\frac{|\Delta_{n,n+p}({\bf q})|^2}
{p^2\omega_c^2}~~.
\label{eix}
\end{equation}
$\tilde\varepsilon_{nn}$ and $\tilde v_{Fn}$ determine ``quasiparticle"
density of states (DOS) at the FS and their
${\bf q}$-dependence describes the {\em ``broadening"}
of LLs by the potential scattering induced
through the spatial non-uniformity of $\Delta ({\bf r})$.
This effect is {\em secondary} to the leading pairing-induced
{\em ``splitting"} of LLs in Eq. (\ref{eii}).
Finally, the renormalized paring matrix element 
at the FS is:
$$\tilde\Delta_{nn}({\bf q})
= \Delta_{nn}({\bf q}) -$$
\begin{equation}
-\sum_{p\not=0,p'\not=0}
\frac{\Delta_{n,n+p}({\bf q})\Delta^*_{n+p,n+p'}(-{\bf q})
\Delta_{n+p',n}({\bf q})}{pp'\omega_c^2}~~.
\label{eixp}
\end{equation}

This ``quasiparticle" construction is not an ordinary
perturbation theory. It keeps the ``normal" and ``paring"
self-energies separate by postponing the Nambu rotation until
the last step. It can be viewed as a {\em ``quasiparticle"
representation of the BSC quasiparticle spectrum}.
Such representation is valid as long as we are interested
in excitations near the FS. 

The above construction can be carried out to any desired
degree of accuracy in $\Delta/\omega_c$, although the
amount of algebra rapidly increases. Certain
general statements, however, can be made. In particular,
the excitation spectrum determined by the BdG equations
(\ref{ei}) is {\em gapless} in a {\em finite} interval
$\Delta/\omega_c\in [0,x_c]$, where $x_c\not= 0$ is
some critical value. This property
can be demonstrated by noticing first that all $\Delta _{nm}({\bf q})$
have a set of common zeroes whose location in the
${\bf q}$-space depends only on whether $n+m$ is
even or odd. These are so-called Eilenberger zeroes,
$\{ {\bf q}^E_i\}$.\cite{sasa}
In the leading order (\ref{eii}) the gap is determined by
the diagonal matrix elements, $\Delta _{nn}({\bf q})$, 
and it vanishes at ${\bf q}= {\bf q}^E_i$
as well as at
numerous other points $\{ {\bf q}^{nn}_j\}$
within the magnetic Brillouin
zone (MBZ). As the matrix elements coupling
different $\psi_{n,p}$'s are turned on,
it is clear from (\ref{ev}) that 
$D_{nn}({\bf q}={\bf q}^E_i)$ involves only
terms with an {\em odd} number of matrix elements
$\Delta _{n+p,n+p'}({\bf q})$, i.e.
\begin{equation}
D_{nn} \sim \Delta_{n,n+p}
\Delta_{n+p,n+p_1}\Delta_{n+p_1,n+p_2}\cdots\Delta_{n+p_k,n}~~.
\label{ex}
\end{equation}
Note now that it is impossible to form such a string (\ref{ex})
of odd number of $\Delta$'s without at least one 
$\Delta_{n+p',n+p''}({\bf q})$ having 
$n+p'+n+p''$ {\em even}. But such a matrix element also must
vanish at same $\{ {\bf q}^E_i\}$ as
$\Delta _{nn}({\bf q})$. Consequently, at these points,
${\bf q}={\bf q}^E_i$,
the spectrum {\em remains gapless}. This statement 
is correct to {\em all}
orders in perturbation theory and therefore is exact
as long as perturbative expansion itself is well-defined.
We expect that the radius of convergence
of the perturbative expansion extends from $\Delta/\omega_c
= 0$ to $\Delta/\omega_c =x_c \sim 1$.\cite{sasa,footv} 
At this critical point, the
perturbation theory breaks down due to band crossings
between neighboring LL branches. For $\Delta/\omega_c > x_c$
the gaps will open up at $\{ {\bf q}^E_i\}$ signaling
the destruction of the LL structure by strong mixing
due to increasing $\Delta$. Ultimately, for $\Delta/\omega_c \gg x_c$,
there is a crossover to the low-$H$ regime of mini-gapped
states in well-separated vortex cores\cite{sasa},
as established by Norman {\it et al.}\cite{norman}.
This phenomenon of band crossings
and disappearance of gapless excitations has
been already noticed in numerical solutions of the
BdG equations.\cite{sasa} Here we elucidate its
physical origin through analytic arguments.

%pedro
We now compare our analytic results with the numerical solution
of the BdG equations (\ref{ei}).
Such solution is very time
consuming in the 3D case so we have solved (\ref{ei}) for
the 2D case.
Our analytic results can be adapted to the 2D case but with
one key {\em difference}: in a 2D system 
the FS is a {\em line} not a {\em sheet}. 
Consequently, the gapless and ``near" gapless points in the
MBZ might not be located on the FS.
In contrast, the Fermi ``sheet" of a LL quantized 3D material
contains {\em all} of the MBZ. As a result, the gapless
character of the excitation spectrum will be less
pronounced in 2D systems even though the effects of LL
quantization itself are more in evidence there than in 
quasi 3D materials. This difference between 3D and 2D systems
should lead to observable effects 
in dHvA oscillations\cite{sasa,norman,maniv},
thermodynamics and transport.

We obtain the spectrum as a function of $\Delta$ for
different $n_c\equiv\mu/\hbar \omega_c$.
The eigenvalues are grouped in pairs centered around the LL
energies.
As $\Delta$ grows, the bands start to broaden until eventually the gap between
them goes to zero and the band-crossing occurs. The sequential eigenvalues
have a mirror-like shape with the neighboring bands.
We fixed $\mu$ such that $\{ {\bf q}^E_i\}$ points
remain gapless.
In 2D, if $\mu$ is adjusted so that an originally gapless point in
the MBZ (\ref{eii}) is on the FS, then the gap opened
by the off-diagonal terms should be due to the pairing term and thus
$\sim \Delta^3$. In general, however, 
the normal contributions at ${\bf q}=\{ {\bf q}^{nn}_j\}$
gapless points are finite and their
gaps will be $\sim\Delta^2$.
A fit of the normal and pairing self-energy
contributions for ${\bf q}={\bf q}^{nn}_j$
as a function of $\mu$, shows that there
is a value of $\mu$ for which the coefficient of the $\Delta^2$ term
vanishes and therefore close to this point 
the $\Delta^3$ term prevails.
This provides a criterion for locating the ``renormalized" 3D FS 
at all $\{ {\bf q}^{E}_{i},{\bf q}^{nn}_j\}$.\cite{footv}

An effective measure of the importance of low energy 
excitations is the density of states (DOS). 
The shape of the DOS at low energies
is determined by the first and second order zeroes 
found in the spectrum.\cite{footvi}
The exact DOS for small $\Delta$ and DOS within the DA 
are basically equal since $\Delta/\omega_c\ll 1$. There
is therefore a large DOS at low energies. 
As $\Delta$ grows a gap appears in 2D. Yet, if
$\mu$ is set to emulate 3D, the DOS remains similar to the DA.
As $n_c$ grows, the region over which the DOS spreads
becomes a smaller fraction of $\Delta$, since the number of zeroes grows
with $n_c$ and the effect of the off-diagonal perturbations decreases, 
likely due to large cancellations in (\ref{evii}).
In Fig. 1 we show the DOS for $n_c=40$ and
$N=10$\cite{sasa,norman} for the values of $\Delta=0.01, 0.5, 4$. The 
energy is rescaled by $\omega_c$. The
nearly diagonal case, $\Delta=0.01$, and $\Delta=0.5$ appear very similar.
They both show a linear dependence in the DOS
indicating qualitatively same behavior. 
The case $\Delta=4$ is qualitatively different.
A gap is appearing at low energies and the next band has a small
gap to the lowest one indicating the band-crossing is 
imminent.
We can see that as $\Delta$ grows some structure is detected in the
shape of the DOS that clearly develops for the $\Delta=4$ case.
These results demonstrate that within our pocket ($H>H^*, T<T^*$) the
DOS is qualitatively similar to the DA
and gapless and nearly gapless points 
$\{ {\bf q}^{E}_{i},{\bf q}^{nn}_j\}$
still dominate low energy properties.

Our results have direct and observable consequences for a broad range of thermodynamic
and dynamic properties. Specific heat, ultrasound attenuation,
STM and other forms of tunneling, and many other phenomena, will
exhibit power-law-like behaviors, characteristic of low-energy 
BCS quasiparticles.  All these diagnostic tools can now be used to
explore the ``extreme" pocket, identify gapless 
behavior and test our prediction 
that the high- and the low-field regime of type-II superconductors
differ in this essential respect.

%kraj

ZT has benefited from a 
discussion with Prof. J. R. Schrieffer
which served as a significant motivation for this work.
The authors thank Prof. S. Dukan for discussions.
This research was supported in part by the NSF
grant DMR-9415549. PDS acknowledges partial support in the
form of a PRAXIS XXI Fellowship.

%%%%%%%%%%%%%%%%%%%%%%%%%%%%%%%%%%%%%%%%%%%%%%%%%%%%%%%%%%%%%%%%%%%%%%%%%%%%%%

\newpage

\begin{figure}[t]
\epsfxsize=7.1cm
\epsffile{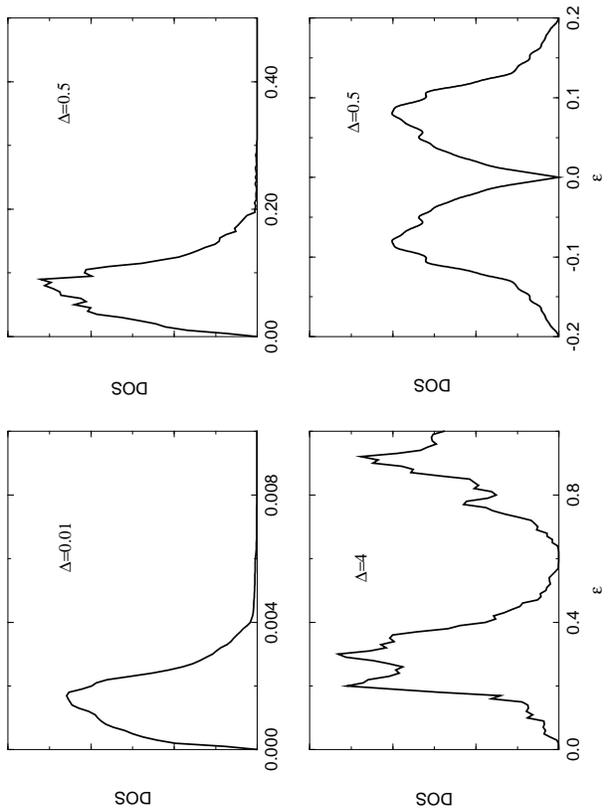}
\vskip 0.3in
\caption[]{DOS (in arbitrary units) for several 
values of $\Delta$ (see text). The inset shows the
``extreme" pocket (shaded area).
}
\label{fig1}
\end{figure}


\begin{references}
\bibitem{footi} Note that,
for the effective mass, $m^*$,  of order of the 
electron mass, $\hbar\omega_c (H=$1 Tesla) $\sim$ 1.34 Kelvin.
\bibitem{footv} The value of $\Delta$ for 
which the band-crossing occurs increases with
the value of $n_c$, since 
the relevant energy scale goes as 
$\sim\sqrt{2} \Delta /(2 \pi n_c)^{1/4}$.\cite{sasa,norman} 
This implies that, for large $n_c$, the range of values
of $\Delta$ for which the perturbation expansion is adequate is larger.
\end{references}
\end{document}